\documentclass[
 aps, pra,
 amsmath,amssymb,
 11pt,
 final,
tightenlines,
 twoside,
 twocolumn,
 nofloats,
nofootinbib,
 superscriptaddress,
showkeys,
showkeywords,
 ]
{revtex4-2}

\usepackage[T1]{fontenc}
\usepackage[utf8x]{inputenc}
\usepackage[english]{babel}
\usepackage{graphicx}
\usepackage{dcolumn}
\usepackage{bm}
\usepackage{longtable}

\input{maik.rty}

\setcitestyle{authoryear,round}
\def\saoname{Special Astrophysical Observatory,  Russian Academy of Sciences,
              Nizhnii Arkhyz, 369167 Russia}

\def\squareforqed{\hbox{\rlap{$\sqcap$}$\sqcup$}}

\def\sq{\ifmmode\squareforqed\else{\unskip\nobreak\hfil
\penalty50\hskip1em\null\nobreak\hfil\squareforqed
\parfillskip=0pt\finalhyphendemerits=0\endgraf}\fi}

\def\utw{\smash{\rlap{\lower5pt\hbox{$\sim$}}}}

\def\udtw{\smash{\rlap{\lower6pt\hbox{$\approx$}}}}

\def\diameter{{\ifmmode\mathchoice
{\ooalign{\hfil\hbox{$\displaystyle/$}\hfil\crcr
{\hbox{$\displaystyle\mathchar"20D$}}}}
{\ooalign{\hfil\hbox{$\textstyle/$}\hfil\crcr
{\hbox{$\textstyle\mathchar"20D$}}}}
{\ooalign{\hfil\hbox{$\scriptstyle/$}\hfil\crcr
{\hbox{$\scriptstyle\mathchar"20D$}}}}
{\ooalign{\hfil\hbox{$\scriptscriptstyle/$}\hfil\crcr
{\hbox{$\scriptscriptstyle\mathchar"20D$}}}}
\else{\ooalign{\hfil/\hfil\crcr\mathhexbox20D}}%
\fi}}

\begin{document}

\selectlanguage{english}

\keywords{galaxies: gaint radio galaxies}

\title{A sample of giant radio sources from the NVGRC catalog}

\author{\firstname{O.}~\surname{Zhelenkova}}
 \affiliation{\saoname}
  \email{zhe@sao.ru}

\author{\firstname{M.}~\surname{Khoruzhenko}}
\affiliation{Southern Federal University, Rostov-on-Don, 344090 Russia}

\begin{abstract}

The NVGRC catalog comprises radio sources selected by a pattern recognition algorithm as candidates for giant radio sources (GRSs). In addition to GRSs, the catalog includes sources with projected sizes smaller than 0.7~Mpc, as well as sources whose components are themselves distinct radio sources but were mistakenly merged into a single object.

We inspected 370 NVGRC sources within the range $00^{h}00^{m} < RAJ < 05^{h}20^{m}$, along with additional radio sources located within one degree neighborhood surrounding each target.

In the examined sample
48\% of objects were classified as giant radio sources,
14\% as sources with projected linear sizes below 0.7~Mpc,
38\% as physically unrelated sources erroneously grouped by the recognition algorithm.
We identified a total of 197 GRSs, including 72 previously known giant radio galaxies (GRGs) or giant radio quasars (GRQs), and 125 newly confirmed GRSs.

Analyzing the distribution of FRI-type giants across four redshift bins, we found that at $z < 0.05$, the proportions of FRI and FRII sources are approximately equal. However, at $z > 0.15$, the fraction of FRI giants drops sharply. The dominance of FRII giants in GRS lists is likely a result of observational selection effects imposed by the sensitivity limits of current radio surveys.

According to the NVSS and VLASS cutouts, 33\% of the sources show signs of fading; 25\% show signs of resuming radio activity; and 38\% have curved, deformed radio lobes. A combination of these features is observed in some radio sources.
The sample includes of 74\% galaxies, 
15\% IR-excess galaxies, which—based on WISE photometric data—are likely quasars,
and 11\% confirmed quasars.

Visual inspection of optical survey cutouts revealed that many host galaxies have close neighbors or belong to known groups or clusters. So, 
39\% of radio sources have neighboring galaxies within 50~kpc, and
28\% are members of galaxy groups or clusters.
Thus, approximately 70\% of GRSs reside in relatively dense environments, and this fraction may be even higher.

\end{abstract}

\maketitle

\section{INTRODUCTION}

Giant radio sources (GRSs) are galaxies or quasars whose radio structures, projected onto the celestial plane, exceed a linear size of 0.7\,Mpc. 
Some of the largest known GRSs reach sizes of approximately 5\,Mpc, again comparable to cluster scales. 

By 2020, around 900 giant radio galaxies (GRGs) had been identified~\citep{1974Natur.250..625W, 2001A&A...370..409L, 2001A&A...374..861S, 2001A&A...371..445M, 2006A&A...454...85M, 2005AJ....130..896S, 2014AstBu..69..141S, 2017MNRAS.469.2886D, 2020A&A...642A.153D, 2018ApJS..238....9K}, and they were considered relatively rare objects. 
As of now, more than 11,500 GRSs have been cataloged~\citep{2021Galax...9...99A, 2023A&A...672A.163O, 2024A&A...691A.185M}. 
Their abundance is particularly notable in the coverage area of the LoTSS low-frequency survey, which benefits from high sensitivity. In other regions of the sky, the detection rate remains relatively low due to the absence of similarly sensitive low-frequency observations.

Several hypotheses have been proposed to explain the large sizes of GRGs:
\begin{enumerate}
    \item \textbf{Environmental factors:} The radio source may reside in a low-density intergalactic medium (IGM), allowing its lobes to expand with minimal resistance~\citep{2008ASPC..395..380S, 2009MNRAS.393....2S, 2015MNRAS.449..955M}.
    \item \textbf{Evolutionary age:} GRSs may represent old radio sources whose extended structures have grown over time~\citep{1997MNRAS.292..723K}.
    \item \textbf{Central engine properties:} The size may be governed by intrinsic characteristics of the galaxy nucleus, such as black hole mass, spin, and accretion rate~\citep{2012MNRAS.426..851K}.
\end{enumerate}

It is widely believed that GRSs represent the final evolutionary stage of radio sources powered by active galactic nuclei. Studies by \citet{1999A&A...345..769M, 1999A&A...344....7P, 2003PASA...20...19M, 2008MNRAS.385.1286J} have shown a tendency for spectral age to correlate with linear size, suggesting that older sources tend to be larger. Nevertheless, compact radio sources with ages up to $10^8$ years have also been observed~\citep{2011A&A...526A.148M}.

Most known giant radio sources (GRSs) are located at low redshifts and are associated with bright elliptical galaxies. These galaxies typically host FRII-type radio sources~\citep{1974MNRAS.167P..31F}, characterized by radio luminosities in the range $10^{23}$–$10^{28}$\,W$\cdot$Hz$^{-1}$ at 1.4\,GHz.

GRSs serve as valuable probes of the intergalactic medium (IGM). Their extended radio lobes interact with the surrounding environment, often producing structural asymmetries that reveal properties of the IGM. Due to their large physical size, GRSs also enable the study of the distribution of the warm-hot intergalactic medium within the large-scale structure of the Universe~\citep{2012MNRAS.426..758P, 2015MNRAS.449..955M, 2015aska.confE.109P, 2009MNRAS.393....2S}.

GRSs transport matter from the host galaxy over megaparsec scales, enriching the IGM and interstellar medium with non-thermal particles and magnetic fields~\citep{1994RPPh...57..325K}. This magnetized plasma can persist for billions of years and may serve as a source of high-energy particle injection into the intracluster medium~\citep{2001ASPC..250..454E, 2010Sci...330..347V}. GRGs are thus considered important agents in the magnetization of the IGM~\citep{2022A&A...660A...2O}.

The radio lobes of GRSs, extending over megaparsec scales, are the largest natural reservoirs of magnetic fields and non-thermal relativistic particles associated with galactic systems. These lobes retain most of the energy released by central black holes over long timescales \citep{2001ApJ...560..178K}. 
Consequently, GRSs provide a useful means of estimating the total energy output of active galactic nuclei. Moreover, the extended lobes, filled with charged particles, are sufficiently large to accelerate particles to ultra-high energies. It is hypothesized that shock waves within radio jets and GRS lobes may contribute to the generation of cosmic rays~\citep{2004ApJ...604L..77K, 2009MNRAS.393.1041H}.

Studies have shown that the density of the IGM is relatively low in the vicinity of some GRSs~\citep{2006A&A...454...85M, 2015MNRAS.449..955M}. However, no significant association has been found between GRSs and cosmic voids~\citep{2018ApJS..238....9K}. Furthermore, \citet{2009ARep...53.1086K} demonstrated that there is no correlation between the linear size of radio sources and the density of galaxies in their surroundings.

Giant radio galaxies with angular sizes exceeding $4^{\prime}$ are of particular interest for separating radio source emission from the cosmic microwave background. They also contribute to the angular power spectrum, which plays a role in constraining cosmological models~\citep{2014ARep...58..506S, 2016AstBu..71..139V}.

Since the IGM density increases with redshift as $\rho \propto (1+z)^3$~\citep{1989AJ.....97....1K}, the expansion of radio lobes becomes increasingly difficult at higher redshifts. Additionally, the surface brightness of radio structures decreases with redshift as $(1+z)^{-4}$, which complicates the detection of extended GRG lobes in earlier cosmological epochs.

For this study, we utilized the NVGRC catalog compiled by \citet{2016ApJS..224...18P}, which presents a list of candidate giant radio sources with angular sizes $\geq 4^{\prime\prime}$, selected from the NVSS catalog using pattern recognition algorithms. In that work, the identification of host galaxies and the determination of their redshifts necessary for calculating the projected linear sizes of radio sources—were not performed.

We conducted a visual inspection of 370 NVGRC objects, representing approximately 25\% of the catalog. To refine the radio morphology and identify host galaxies, we employed all publicly available radio, optical, and infrared surveys.

A similar study was carried out by \citet{2017MNRAS.469.2886D}, who considered only those NVGRC objects for which a radio core was detected in VLASS maps. In contrast, our analysis includes all NVGRC objects, regardless of whether a radio core was identified in VLASS data.

Throughout this paper, we adopt a flat $\Lambda$CDM cosmology based on Planck results: $H_0 = 67.4\,\mathrm{km}\cdot\mathrm{s}^{-1}\mathrm{Mpc}^{-1}$ and $\Omega_m = 0.315$~\citep{2020A&A...641A...6P}. The spectral index $\alpha$ of a radio source is defined by the relation $S_{\nu} \propto \nu^{\alpha}$.

\section{Search for giant radio sources in the NVGRC catalog}

The NVGRC catalog~\citep{2016ApJS..224...18P} is derived from the NVSS catalog~\citep{1998AJ....115.1693C}, in which a single radio source may be represented by multiple entries due to its extended structure.
To identify giant radio sources, \citet{2016ApJS..224...18P} employed the Oblique Classifier One (OC1) software, which implements a decision tree algorithm~\citep{Murthy1994}. The OC1 classifiers were trained on a reference set of 48 GRGs characterized in \citet{2001A&A...370..409L}, using their morphological and photometric properties.
As a result of this classification procedure, a list of 1616 candidate GRSs was compiled.

\subsection{Radio and optical identification of candidates}

Due to the large angular sizes of GRSs, identifying their host galaxies is a non-trivial task. Moreover, when the surface brightness of the radio lobes is low, the recognition of the GRS itself becomes challenging. If a candidate exhibits a radio core that coincides with an optical object, the identification is straightforward and reliable.

For radio sources without a detectable radio core in VLASS maps, it is first necessary to identify the radio morphology and then the probable position of the host galaxy.
Recent radio surveys at low and high frequencies help classify radio structures of sources.

Optical or ultraviolet radiation obscured by dust structures surrounding the accretion disk of an active galactic nucleus is re-emitted in the mid-infrared range. And mid-IR data can be useful for identifying the host galaxy.

\citet{2020A&A...642A.153D} conducted a large-scale search for GRSs among NVGRC objects with a radio core detected in VLASS maps. In contrast, our study considered all NVGRC candidates, including those without a VLASS-detected core. Furthermore, we examined one-square-degree neighborhoods around each candidate.

To work with multiple catalogs and surveys, we used the Aladin Sky Atlas~\citep{2000A&AS..143...33B} and TOPCAT~\citep{2005ASPC..347...29T} software tools.

To determine the radio morphology of NVGRC objects, we utilized data from
NVSS, VLASS~\citep{2021ApJS..255...30G}, and FIRST~\citep{2015ApJ...801...26H},
TGSS~\citep{2017A&A...598A..78I}, GLEAM~\citep{2017MNRAS.464.1146H}, WENSS~\citep{Rengelink1997}, SUMSS~\citep{2003MNRAS.342.1117M}, RACS~\citep{2020PASA...37...48M}, and
in selected cases, GB6~\citep{1996ApJS..103..427G} and Apertif~\citep{2011JApA...32..557R, 2022A&A...667A..38A}.

Once the radio structure was established, we proceeded with optical identification using 
SDSS~\citep{2020ApJS..249....3A}, PanSTARRS~\citep{2016arXiv161205560C}, DES~\citep{2018ApJS..239...18A}, and Legacy Surveys~\citep{2019AJ....157..168D}.

For sources without a VLASS-detected core, we also examined mid-IR images from WISE~\citep{2013wise.rept....1C}. In cases where two nearby optical objects were plausible host candidates, we used UKIDSS cutouts~\citep{2007MNRAS.379.1599L, 2008MNRAS.391..136L}, selecting the brighter object in the $K$-band as the more likely host. Additionally, we verified the proper motion of candidates using the GAIA catalog~\citep{2018A&A...616A...1G} to exclude stellar contaminants.

Spectroscopic or photometric redshifts were retrieved from the Simbad~\citep{2000A&AS..143....9W}, NED~\citep{1995ASSL..203...95H}, NOIR DataLab~\citep{2019BAAS...51g..61O}, and Vizier~\citep{2000A&AS..143...23O} databases.

\subsection{Measurement of angular sizes}

For FRII-type sources, the angular size is typically measured as the distance between the hot spots. For FRI and hybrid FRI/FRII sources, it is estimated as the distance between the outer edges of the radio lobes. In the case of curved sources, the angular size is measured along the so-called ridge of the radio structure.

\citet{2016ApJS..224...18P} measured angular sizes along the outer lobe edges at a level of $3\sigma$ above the background for all morphological types (FRI, FRII, and hybrids). In our study, we followed the same approach and did not differentiate between FRI and FRII classifications. All measurements were performed using the ``distance'' tool in the Aladin Sky Atlas.

For sources with curved lobes, angular size estimation inevitably involves a degree of subjectivity.

We measured the angular sizes of radio sources using both NVSS and VLASS cutouts. Some sources in the VLASS maps show only a compact core without any noticeable lobes. In these cases, we did not measure the sizes from the VLASS data, and there were 29 such sources.

We compared the projected sizes of GRSs from our sample with those reported in \citet{2018ApJS..238....9K}, \citet{2020A&A...642A.153D}, and \citet{2023A&A...672A.163O}. 

The results of this comparison are summarized in Table~\ref{tab:Tab1}. The first column lists the GRG catalogs being compared, using the following abbreviations: D20 — \citet{2020A&A...642A.153D}, K18 — \citet{2018ApJS..238....9K}, O23 — \citet{2023A&A...672A.163O}, and OL — our list. The second column indicates the number of sources matched between the respective catalogs. The third column presents the mean difference and root mean square (RMS) deviation in projected size (in Mpc).  
\begin{table}
	\centering
	\caption{The difference in the measured LAS of the giants.\\}
	\label{tab:Tab1}
	\begin{tabular}{lrc} 
		\hline
		List    & N, obj. & $\Delta D \pm RMS$ \\
		\hline
		K18-D20 & 257 & 0.05$\pm$0.35 \\
		O23-K18 & 19  & 0.20$\pm$0.28 \\
            \hline
            OL-O23  & 7   & 0.18$\pm$0.21 \\
            OL-D20  & 51  & 0.29$\pm$0.63 \\
            OL-K18  & 24  & 0.29$\pm$0.18 \\
            \hline
	\end{tabular}
\end{table}

A systematic offset of 0.2–0.3\,Mpc was found between our measurements and those reported in \citet{2018ApJS..238....9K}\footnote{
Our measurements of the angular size of NVGRC\,J005748.3+302114 differ from \citet{2018ApJS..238....9K}.
In our opinion, the northern component of the source appears more extended.}, \citet{2020A&A...642A.153D}\footnote{
Discrepancies in angular size for two sources arise from differing redshift values. For NVGRC\,J000622.1+263549, we used $z_{\mathrm{ph}} = 0.835$ from the DESI survey instead of $z_{ph} = 0.436$. For NVGRC\,J042220.9+151101, we adopted $z_{sp} = 0.072$ from NED rather than $z = 0.409$.}, and \citet{2023A&A...672A.163O}. 
This difference is most likely due to our methodology: we measured the angular extent from the outer edges of the radio lobes rather than between the hot spots.

\section{Results of a visual inspection of GRS-candidates}

Of the 1616 objects listed in the NVGRC catalog, we examined 370 sources (23\%) within the right ascension range $00^{\mathrm{h}}00^{\mathrm{m}} < \mathrm{R.A.} < 05^{\mathrm{h}}20^{\mathrm{m}}$. In addition, we inspected radio sources with angular sizes $\geq 2.5^{\prime}$ that appeared within $1\Box^{\circ}$ NVSS cutouts centered on each NVGRC object.

Some NVGRC entries consist of NVSS components that correspond to physically unrelated radio sources. In certain cases, only one component of an NVGRC candidate was found to belong to a radio source that we classified as a GRS. In other instances, we identified GRSs within the NVSS cutouts that were not included in the NVGRC catalog. Accounting for these findings, we report 20 additional GRSs that were not listed in the original NVGRC catalog.

It is worth noting that the source J003419.3+011857 is treated as a group of sources in \citet{2011ApJS..194...31P}, while \citet{2016ApJS..224...18P} classify it as a GRS candidate. This object consists of two relatively close radio quasars, as confirmed by their SDSS DR16 spectroscopic redshifts. The southern source in this pair was classified by \citet{2012MNRAS.426..851K} as having a projected linear size below 0.7\,Mpc. However, based on our measurements and a redshift value higher than that used in \citet{2012MNRAS.426..851K}, we find that its projected size exceeds 0.7\,Mpc.

Another candidate, NVGRC J035339.2-011319, was classified as a GRG by \citet{2020A&A...642A.153D}. However, based on VLASS maps, the radio morphology of its northern and southern components is more consistent with two separate double-lobed radio sources. The northern component has a clear optical counterpart in the PanSTARRS survey, while the southern component corresponds to a faint optical object visible in DECals cutouts. Therefore, we did not include this source in our list of giant radio galaxies.

In total, we identified 197 GRSs in our sample, distributed as follows:
\begin{itemize}
    \item \textbf{GRGs with spectroscopic redshifts:} 86 sources, including 50 previously known GRGs and 36 newly confermed GRSs. Among these, 6 are not listed in the NVGRC catalog.
    
    \item \textbf{GRGs with photometric redshifts:} 72 sources, of which 17 are included in \citet{2020A&A...642A.153D}, and 55 were discovered in this work. Five of these are not present in the NVGRC catalog. For two of these five, host identification remains uncertain due to complex radio morphology.
    
    \item \textbf{GRQs with spectroscopic redshifts:} 8 sources, including 5 previously known GRQs and 3 newly discovered ones. Two of these are not listed in the NVGRC catalog.
    
    \item \textbf{Quasars with photometric redshifts:} 8 sources, all newly discovered in this study. Four of them are not included in the NVGRC catalog.
    
    \item \textbf{Sources without redshift information:} 23 parent objects lack redshift data. We estimated their redshifts using the magnitude-redshift relationship.
\end{itemize}

We were unable to confidently determine the radio morphology of seven NVGRC objects: J000106.4+340303, J005451.5+564842, J021329.0+292139(2)\footnote{
Here and below, the number in parentheses indicates the components of the NVGRC object. The southern component is marked with 1, and the northern component with 2.
}, J025347.1$-$200007, J032145.1+514855, J035800.3$-$393629(2), and J050341.2$-$191142.

Three NVGRC sources J011352.3+622434, J043503.2+215527, and J051219.4+131945 have very faint optical hosts. These hosts are visible only in PanSTARRS cutouts but are absent from the PanSTARRS catalog. Their positions align well with the expected centers of the radio structures. Notably, J011352.3+622434 and J051219.4+131945 are listed in the WISE catalog.

Among the 370 NVGRC objects we studied, several radio sources were found that contain two NVGRC objects. For example,
J005748.3+302114 and J010001.3+300249 are a single radio source. This also applies to
J022318.0+425939 and J022251.6+425744,
J050533.7-285707 and J050540.8-282445,
J051601.7+245826 and J051605.7+245833.

Additionally, J024733.6+615632 and J035322.1+355212 were identified as HII regions.

\section{Types of parent objects}

To classify host objects as either quasars or galaxies, we utilized data from the Simbad, NED, Vizier, SDSS, and Legacy Surveys (LS) databases. It should be noted that in Simbad and NED, an optical object may be simultaneously classified as both a galaxy and a quasar. Such dual-type assignments often arise from updated or conflicting information in the literature. Additionally, the host may belong to the class of ``changing-look'' AGNs~\citep{2003MNRAS.342..422M, 2014ApJ...796..134D}.

For sources with only photometric data, we applied the quasar selection criteria based on WISE color indices~\citep{2018ApJ...861...37G, 2022ApJ...934..119G}, using photometry from the AllWISE catalog\footnote{VizieR On-line Data Catalog: II/328}. The adopted criteria were: 
\begin{align*}
0.5 < W1 - W2 < 2, \\
2 < W2 - W3 < 4.5, \\
W3 - W4  > 1.9, \\
\end{align*}
where W1, W2, W3, and W4 correspond to the 3.4, 4.6, 12, and 22\,$\mu$m bands of the Wide-field Infrared Survey Explorer (WISE)~\citep{2010AJ....140.1868W}, respectively.

For the faintest hosts, we visually inspected WISE cutouts. If an object was bright in the W1 and W2 bands but undetected in W3 and W4, we classified it as a galaxy. Conversely, if the object was bright in W3 and W4, we classified it as a quasar.

Based on this approach, we divided the host objects into three categories:
galaxies (74\%), quasars (11\%), and IR-excess galaxies with WISE colors consistent with quasars (15\%).

For comparison, the fraction of galaxies among GRG hosts is reported to be 82\% in \citet{2020A&A...642A.153D} and 80\% in \citet{2018ApJS..238....9K}.

\section{Redshifts and radio power}

Of the 197 radio sources analyzed in this study, 94 host galaxies have spectroscopic redshifts, 80 have photometric redshifts, and 23 lack redshift information.

\subsection{Redshift estimation}
\begin{figure}
\center
{(a)\includegraphics[scale=0.3]{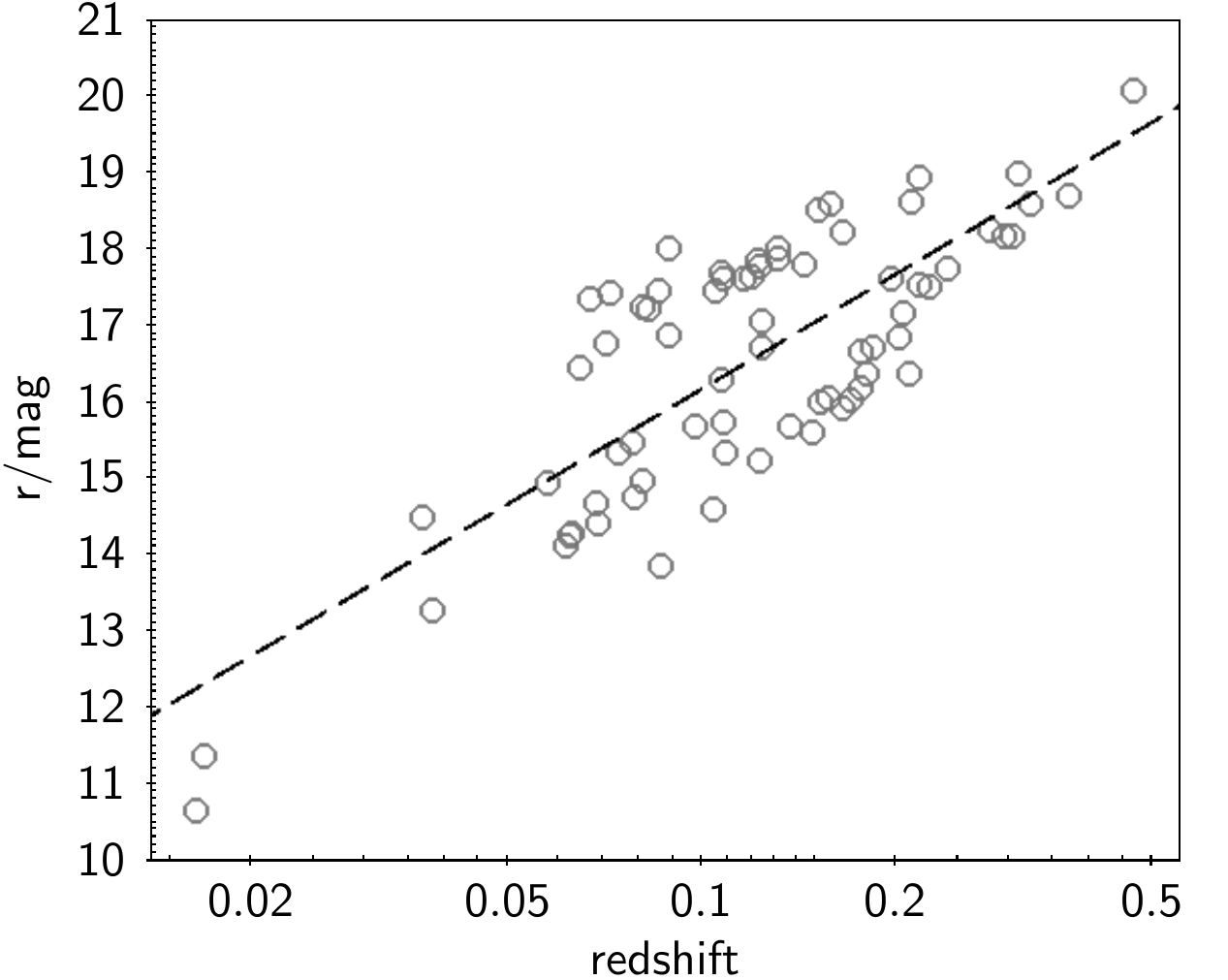}}
{(b)\includegraphics[scale=0.3]{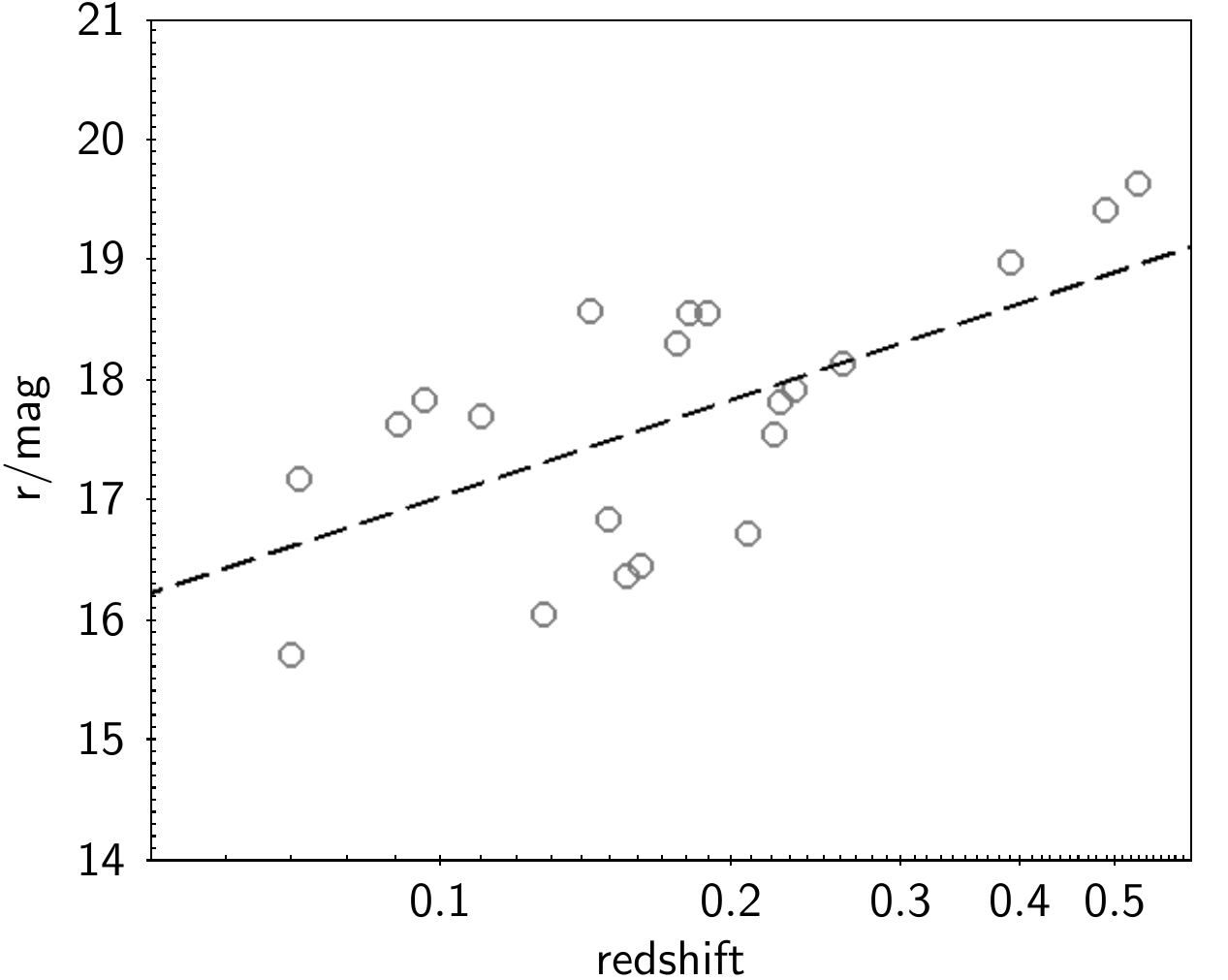}}
\caption{
Scatter plots illustrating the correlation between de-reddened apparent $r$-band magnitudes and spectroscopic redshifts:
(a) shows the distribution for 71 GRGs with known spectroscopic redshifts, along with the linear regression fit (dotted line).
(b) displays the same correlation for 23 GRQs with spectroscopic redshifts.
In both panels, the X-axis (redshift) is plotted on a logarithmic scale.}
\label{fig:Fig1}
\end{figure}
\citet{2018ApJS..238....9K} and \citet{2001A&A...378..826L} reported a correlation between the apparent magnitude of GRG host galaxies and their redshift. This relationship can be used to estimate photometric redshifts for sources lacking direct redshift measurements.

Using $r$-band apparent magnitudes and spectroscopic redshifts of GRGs, we constructed a linear regression model to approximate redshift values. The photometric data and redshift information were obtained from the PanSTARRS, Legacy Surveys (LS), and SDSS catalogs, as well as the NED and SIMBAD databases.

For 71 GRGs with known spectroscopic redshifts, we derived the following empirical relation between the de-reddened $r$-band apparent magnitude ($m_r$) and redshift ($z$), as shown in Fig.~\ref{fig:Fig1}a:
\begin{equation}
\label{form3}
m_r = 5.00 \times \log(z) + 21.16
\end{equation}
This relation yields a Pearson correlation coefficient of $r = 0.78$ and a root mean square (rms) scatter of 1.07\,mag.

For GRQs, we constructed a separate regression using 8 confirmed quasars and 15 galaxies whose WISE color indices suggest quasar-like properties. The resulting relation, shown in Fig.~\ref{fig:Fig1}b, is:
\begin{equation}
\label{form4}
m_r = 2.66 \times \log(z) + 19.69
\end{equation}
with a correlation coefficient of $r = 0.72$ and rms scatter of 0.75\,mag.

When comparing spectroscopic redshifts with those estimated from the above formulas, the rms deviation was approximately 0.07 for galaxies and 0.15 for quasars.

Using these relationships, we estimated the redshift for 23 parent objects (16 galaxies and 7 quasars), including objects fainter than $m_r$=20.6.

The redshift distribution of the studied sample is summarized as follows:
for 94 objects with spectroscopic redshifts, the median redshift is $z = 0.13$;
for 80 objects with photometric redshifts, the median redshift is $z = 0.31$;
ror 23 objects with redshift estimates derived from the empirical magnitude–redshift relations (see Equations~\ref{form3} and~\ref{form4}), the median redshift is $z = 0.63$.

\subsection{Radio loudness}
\begin{figure}
\center
{(a)\includegraphics[scale=0.21]{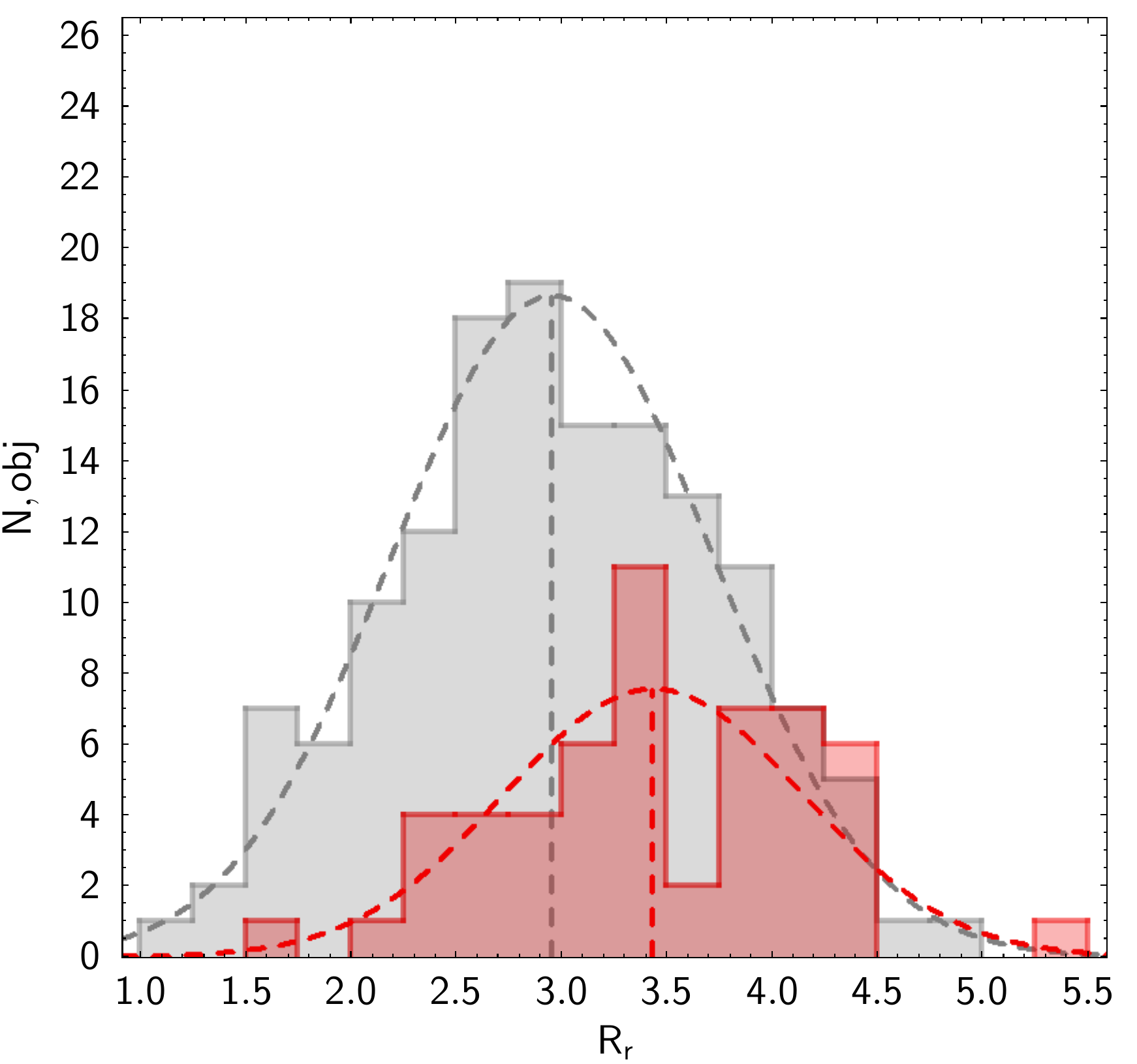}}
{(b)\includegraphics[scale=0.21]{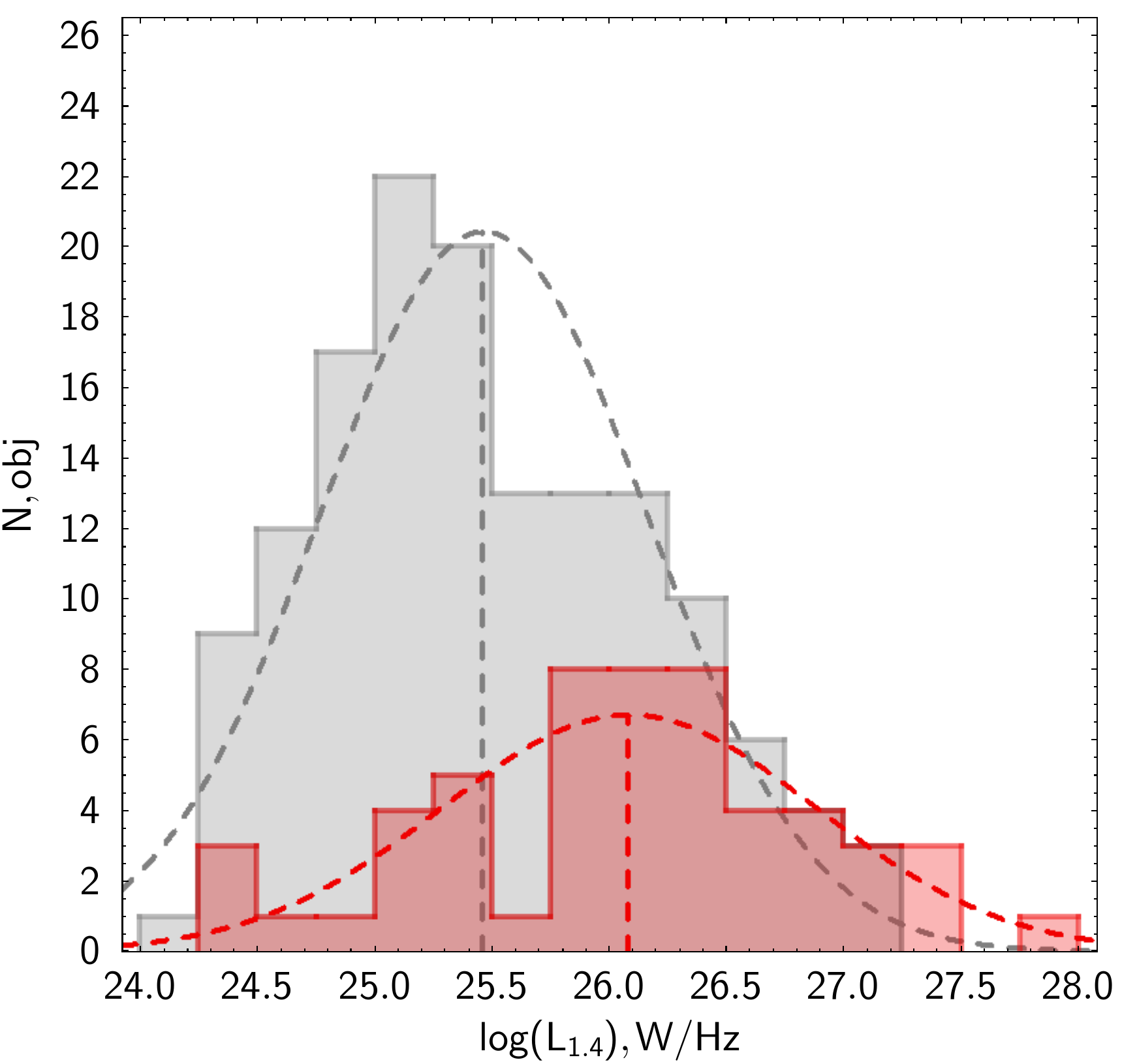}}
\caption{
Histograms and their Gaussian fits illustrating the distributions of two key parameters:
(a) distribution of radio loudness indices;
(b) distribution of radio luminosities at 1.4\,GHz, expressed in W\,Hz$^{-1}$.
In both panels, the data are separated by host type:
galaxies are shown in grey; quasars and IR-excess galaxies are shown in red.}
\label{fig:Fig2}
\end{figure}
Relativistic jets, formed through the extraction of rotational energy from supermassive black holes via magnetic fields and sustained by accreting matter, are highly efficient emitters of radio synchrotron photons. Their presence distinguishes active galactic nuclei (AGN) as members of the radio-loud class. Today, the classification of AGNs into jetted and non-jetted categories has become synonymous with the designation of radio-loud and radio-quiet AGNs~\citep{2019NatAs...3..387P}.

To determine radio loudness, we adopted the approach proposed by~\citet{2002AJ....124.2364I}. Specifically, we use the ratio $R_r$ of radio to optical flux density (without applying a K-correction), calculated according to the formula:
\begin{equation}
\label{eq:form1}
R_r = 0.4 \times (m_r - t_N),
\end{equation}
where $m_r$ is the de-reddened magnitude in the $r$-band, and $t_N$ is the NVSS flux density expressed in AB magnitudes using:
\begin{equation}
\label{eq:form2}
t_N = -2.5 \times \log\left(\frac{F_N}{3631\,\mathrm{Jy}}\right).
\end{equation}

Radio sources with $R_r > 1$ are classified as radio-loud AGNs~\citep{2008AJ....136..684K}.

The calculated $R_r$ values for our sample range from 1.04 to 5.31 (see Fig.~\ref{fig:Fig2}a), indicating that all objects fall within the radio-loud AGN category. For FRI and FRI/II sources, the radio loudness index does not exceed 3.5, whereas FRII sources can exhibit even higher values.

\subsection{Radio Luminosity}

Radio sources are commonly classified by morphological type into FRI and FRII, with the latter being more powerful in the radio domain. The radio luminosities of FRI sources at 1.4\,GHz typically lie in the range of $10^{23}$–$10^{26}$\,W/Hz, while those of FRII sources start from approximately $10^{24.5}$\,W/Hz and extend to higher values~\citep{1994ASPC...54..319O}.

The rest-frame radio power at 1.4\,GHz was estimated using the formula from~\citet{2012MNRAS.426..851K}.

The radio luminosities at 1.4\,GHz for the GRSs in our sample range from $10^{24.2}$ to $10^{27.9}$\,W/Hz (see Fig.~\ref{fig:Fig2}b), indicating that all objects belong to the class of powerful radio sources.

\section{Radio Morphology}

We performed a morphological classification of the giant radio sources (GRSs) using cutouts from the NVSS and RACS surveys, as well as higher-resolution cutouts from the FIRST and VLASS surveys. The sources under consideration consist of between 2 and 20 NVSS components.

Based on NVSS survey maps, we classified 10\% of the sources as FRI type, 3\% as FRI/II type, and 87\% as FRII type. For comparison, the proportion of FRII sources reported by~\citet{2018ApJS..238....9K}, \citet{2020A&A...642A.153D}, and \citet{2021Galax...9...99A} is 90\%, 89\%, and 93\%, respectively.

We compared the ratio of FRI and FRII sources as a function of redshift across three published GRS lists~\citep{2020A&A...642A.153D, 2018ApJS..238....9K, 2023A&A...672A.163O} and our sample. Table~\ref{tab:Tab2} presents statistics on the number of FRI, FRI/II, and FRII sources for four redshift intervals. The list notations follow those used in Table~\ref{tab:Tab1}. In each table cell, the first number indicates the count of FRI and FRI/II sources, while the number after the slash corresponds to FRII sources. The last row of the table shows the average percentage of FRI and FRI/II sources relative to all GRSs within the given redshift interval.

At low redshifts ($z < 0.05$), the number of FRI and FRI/II sources can be comparable to that of FRII sources. However, in the redshift interval $z = 0.15$–$0.20$, the fraction of FRI sources decreases significantly. Due to insufficient data, it is difficult to estimate the fraction of FRI sources at redshifts $z > 0.2$.

We suggest that the low surface brightness of the outer parts of lobes in FRI and FRI/II sources makes GRSs of this type difficult to detect at $z > 0.2$. Consequently, their representation in GRS catalogs is limited.
\begin{table}
	\centering
	\caption{Counts of FRI and FRII sources in redshifts bins}
	\label{tab:Tab2}
	\begin{tabular}{lcccc} 
		\hline
		List & z$<$0.05 & 0.05$\div$0.10 & 0.10$\div$0.15 & 0.15$\div$0.20 \\
	\hline
D20  & 4/7 & 12/29 & 20/43 & 6/37 \\
K18  & 7/5 & 13/35 & 10/34 & 2/32 \\
OL   & 3/1 & 13/13 & 10/26 & 1/25 \\
\hline
 mean & 52\% & 33\%  &  28\%  &  9\% \\
  \hline
\end{tabular}
\end{table}

\subsection{Radio morphological features}

Based on visual inspection of VLASS cutouts, we assigned 10\% of GRSs to sources with core-jet or core-lobe morphology, 17\% to double-lobed sources, 57\% to doubles with a core, and 16\% to triple sources\footnote{We refer to triple sources as those in which the integrated core flux density constitutes 10–20\% of the total source flux in the NVSS, RACS, or TGSS catalogs.}. Thus, 83\% of the sources in VLASS cutouts exhibit a radio core, enabling reliable host galaxy identification.

Deformation or curvature of radio lobes is indicative of the surrounding environment and/or processes occurring near the AGN. Tailed morphologies—WAT (Wide-Angle Tailed) and NAT (Narrow-Angle Tailed)—suggest that the source resides in galaxy clusters or groups~\citep{1976ApJ...205L...1O,2019A&A...626A...8M}. We identified head-tail (HT) features in 22\% of the giant sources.

X-, Z-, and S-shaped radio lobe morphologies are typically attributed to changes in jet orientation, either due to the merger of a smaller galaxy with a massive elliptical host or to instabilities in the accretion disk~\citep{2002MNRAS.330..609D, 2004MNRAS.347.1357L, 2019ApJ...887..266J}.

Sources exhibiting double-double morphology~\citep{2011MNRAS.410..484B} or triple morphology~\citep{2012RAA....12..127G} are classified as AGNs undergoing radiophase restart. We grouped S-, Z-, and X-shaped sources together with double-double and triple sources, as their morphological features suggest activity near the central engine. Such features were identified in 26\% of the sample.

Some GRSs in our sample show combinations of the aforementioned morphological traits. Overall, approximately half of the sources exhibit additional structural features in their radio lobes.

By comparing NVSS and VLASS cutouts, we found that 38\% of the sources display prominent radio lobes at 1.4\,GHz, which are either absent or weakly expressed in the 3\,GHz VLASS maps.

Considering the occurrence of weakly pronounced lobes in VLASS maps, we found that such features appear in 14\% of quasars and IR-excess galaxies, and in 40\% of normal galaxies.

\section{Environment of giant radio sources}
\begin{figure}
\center
{(a)\includegraphics[scale=0.21]{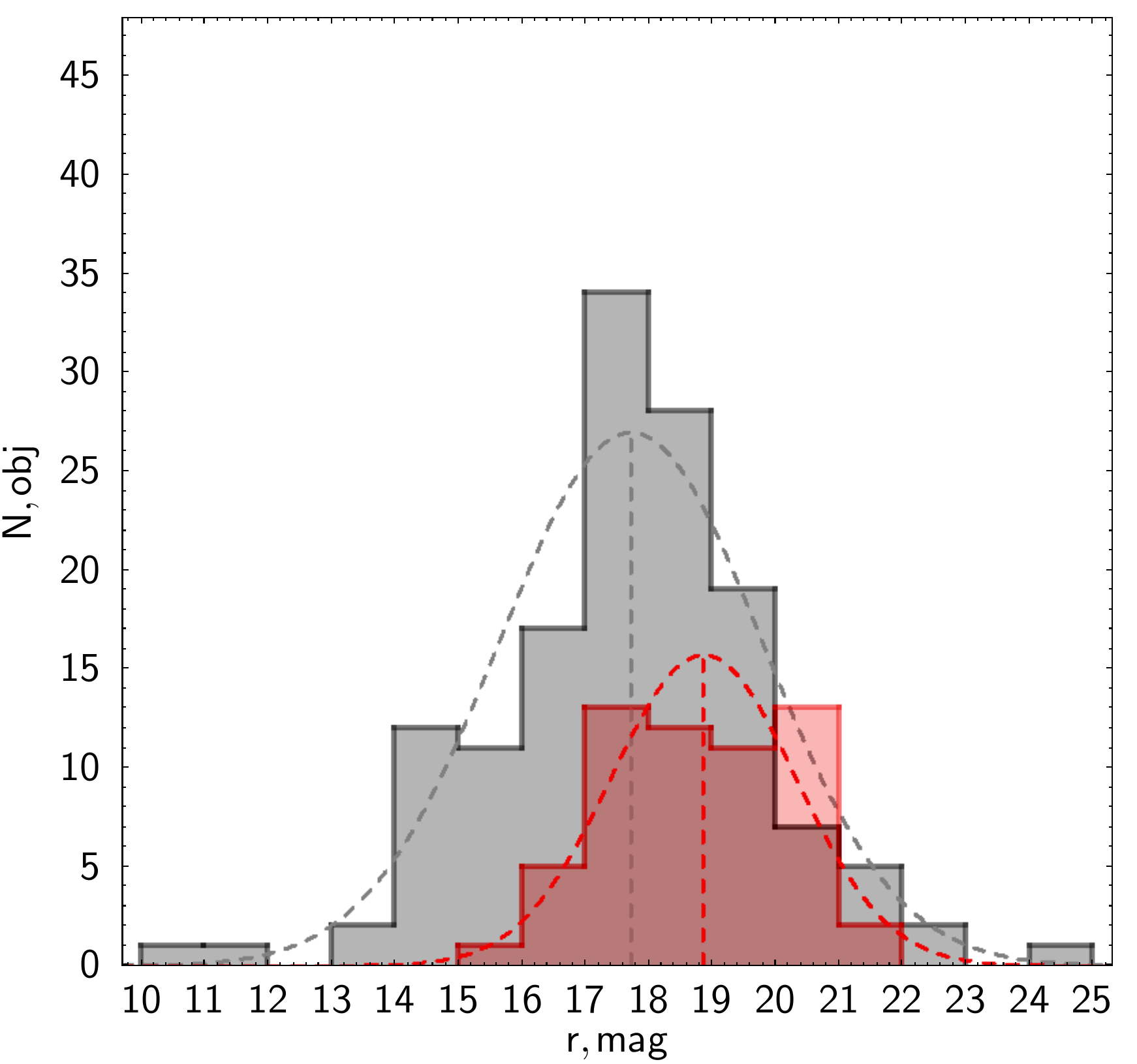}}
{(b)\includegraphics[scale=0.21]{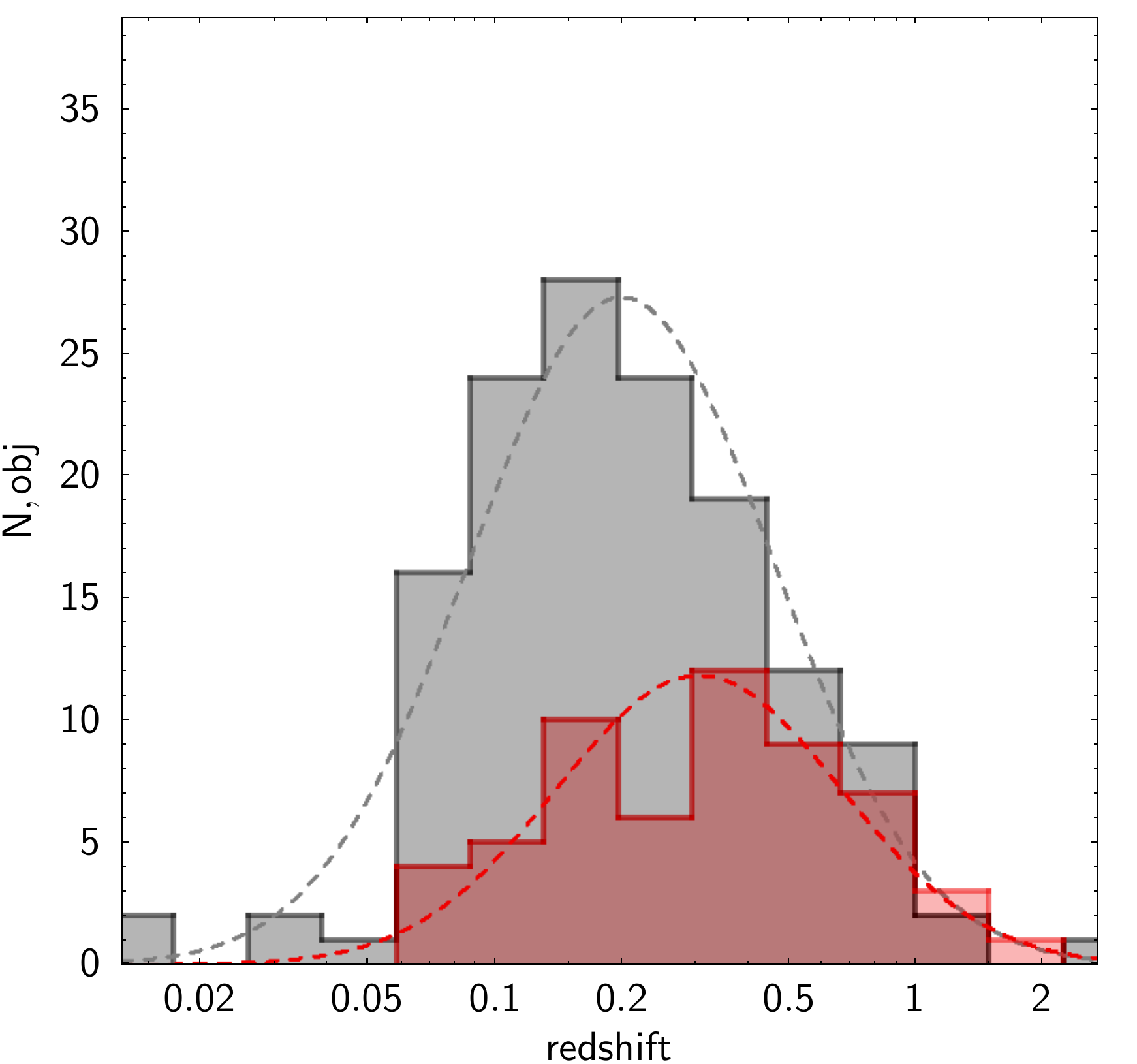}}
\caption{The histograms and their Gaussian fits illustrate the distributions of de-reddened $r$-band magnitudes (panel a) and logarithmic redshifts (panel b) for host galaxies of 140 GRSs with neighbors (gray) and 57 GRSs without neighbors (red).}
\label{fig:Fig3}
\end{figure}

In examining the environments of GRSs, we considered the presence of optical neighbors, deformation of radio lobes, and evidence from the literature indicating host membership in galaxy groups or clusters. During a detailed visual inspection of the surroundings of GRS hosts using optical survey cutouts, we identified neighboring galaxies within projected distances of approximately 50\,kpc, corresponding to angular separations between $1^{\prime\prime}$ and $2^{\prime}$ depending on redshift.

We categorized the GRSs into groups ranging from those with no signs of a nearby environment to those with confirmed group or cluster membership reported in the literature. No neighbors were identified near 36 hosts, and for 21 hosts, redshift information for nearby galaxies was unavailable. The remaining hosts exhibit one or more of the following: neighbors with redshifts consistent with that of the host (within measurement uncertainty), S-, Z-, or X-shaped radio morphology, membership in a known group or cluster, or head-tailed radio morphology.

As a result, we found that 140 sources (71\% of the sample) have close neighbors confirmed by redshift, exhibit morphological features indicative of environmental interaction, and/or are members of galaxy groups or clusters.

Figure~\ref{fig:Fig3} presents histograms of the apparent $r$-band magnitudes (panel a) and logarithmic redshifts (panel b) for GRSs with neighbors (light gray) and those without neighbors (black line).

The median apparent $r$-band magnitudes and redshifts for host galaxies with confirmed neighbors are $18^{\mathrm{m}}.1$ and $z = 0.19$, respectively, while for hosts without neighbors, the corresponding values are $19^{\mathrm{m}}.3$ and $z = 0.31$. Thus, host galaxies without identified neighbors tend to be fainter and more distant than those with confirmed neighbors. We attribute this difference, at least in part, to observational selection effects. It is likely that the true fraction of GRS hosts with neighbors exceeds 70\%.

\section{Summary}

We visually examined objects from the NVGRC catalog within the interval $00^{h}00^{m} < {R.A.} < 05^{h}20^{m}$ to search and confurm for giant radio sources (GRSs). Among the 370 objects inspected, 48\% were classified as GRSs, 14\% had projected sizes smaller than 0.7\,Mpc, and 38\% were independent radio sources erroneously grouped into single systems by the recognition algorithm. Additionally, by inspecting NVSS cutouts of $\sim1\Box^{\circ}$ centered on each NVGRC object, we identified 20 GRSs not included in the catalog. Accounting for these, we estimate the efficiency of the recognition algorithm~\citep{2016ApJS..224...18P} to be approximately 30\%.

Of the 197 GRSs identified in our study, 72 sources (68 galaxies and 5 quasars) are already listed in \cite{2001A&A...370..409L, 2001A&A...374..861S, 2012MNRAS.426..851K, 2018ApJS..238....9K, 2020A&A...642A.153D}. We report 97 newly confurmed GRSs (86 BRKs and 11 quasars) with known spectroscopic or photometric redshifts for their host galaxies. For an additional 23 sources, redshifts were estimated using the empirical $m_r$–$z$ relation.

Morphological classification based on NVSS cutouts revealed that 87\% of the sources are of FRII type. This proportion is consistent with previous studies \citep{2018ApJS..238....9K, 2020A&A...642A.153D, 2021Galax...9...99A}. We analyzed the distribution of FRI and FRII sources across four redshift bins. At low redshifts ($z < 0.05$), the proportions of FRI and FRII sources are roughly equal, but for $z > 0.15$, the fraction of FRI sources declines sharply. This trend likely reflects observational selection effects due to the sensitivity limits of current radio surveys.

According to VLASS maps, 83\% of the sources exhibit core-jet, core-lobe, double-core, or triple morphology, indicating the presence of a radio core and enabling reliable host identification.

By comparing NVSS and VLASS cutouts, we found that 33\% of the sources can be classified as “faded,” 25\% show signs of radiophase restart, and 38\% exhibit deformed radio lobes.

For sources with spectroscopic redshifts, we derived relationships between apparent magnitudes and redshifts, which were then used to estimate redshifts for 23 GRSs lacking direct redshift measurements.

To determine the type of the host galaxies, we primarily used data from the SIMBAD and NED databases. For sources without classification, we applied WISE photometric criteria to distinguish between galaxies and quasars, particularly for faint objects. As a result, our GRS sample comprises 74\% galaxies, 15\% IR-excess galaxies (likely quasars based on WISE photometry), and 11\% confirmed quasars.

Visual inspection of optical survey maps revealed close neighbors and group or cluster membership for many host galaxies, supported by radio morphology and literature data. We found that 140 sources (71\%) reside in relatively dense environments, and this fraction may be even higher.

The total sky area covered by our NVGRC cutouts is approximately $370\Box^{\circ}$. The full sky region within the right ascension range $00^{\mathrm{h}}00^{\mathrm{m}}$ to $05^{\mathrm{h}}20^{\mathrm{m}}$ spans about $8600\Box^{\circ}$. Of the 197 GRSs found in our inspected regions, 20 were missed by the recognition algorithm and are not included in the NVGRC catalog. Extrapolating from this, we estimate that approximately 430 GRSs may have been missed in the full area, implying an algorithmic detection efficiency of roughly 30\%.

\acknowledgments{
This research makes use of services and data provided by the Astro Data Lab at NSF’s NOIRLab. NOIRLab is operated by the Association of Universities for Research in Astronomy (AURA), Inc., under a cooperative agreement with the National Science Foundation.

This work has also made use of the NASA/IPAC Extragalactic Database (NED), which is funded by the National Aeronautics and Space Administration and operated by the California Institute of Technology.

We additionally used the CATS database, available via the Special Astrophysical Observatory website.
}

\bibliographystyle{aspb1}
\bibliography{GRS}

\begin{thebibliography}{83}
\providecommand{\natexlab}[1]{#1}

\bibitem[{Abbott} et~al.(2018)]{2018ApJS..239...18A}
T.~M.~C. {Abbott}, F.~B. {Abdalla}, S.~{Allam}, et~al., \apjs \textbf{239}~(2), 18 (2018).

\bibitem[{Adams} et~al.(2022)]{2022A&A...667A..38A}
E.~A.~K. {Adams}, B.~{Adebahr}, W.~J.~G. {de Blok}, et~al., \aap \textbf{667}, A38 (2022).

\bibitem[{Ahumada} et~al.(2020)]{2020ApJS..249....3A}
R.~{Ahumada}, C.~A. {Prieto}, A.~{Almeida}, et~al., \apjs \textbf{249}~(1), 3 (2020).

\bibitem[{Andernach} et~al.(2021)]{2021Galax...9...99A}
H.~{Andernach}, E.~F. {Jim{\'e}nez-Andrade}, and A.~G. {Willis}, Galaxies \textbf{9}~(4), 99 (2021).

\bibitem[{Bonnarel} et~al.(2000)]{2000A&AS..143...33B}
F.~{Bonnarel}, P.~{Fernique}, O.~{Bienaym{\'e}}, et~al., \aaps \textbf{143}, 33 (2000).

\bibitem[{Brocksopp} et~al.(2011)]{2011MNRAS.410..484B}
C.~{Brocksopp}, C.~R. {Kaiser}, A.~P. {Schoenmakers}, and A.~G. {de Bruyn}, \mnras \textbf{410}~(1), 484 (2011).

\bibitem[{Chambers} et~al.(2016)]{2016arXiv161205560C}
K.~C. {Chambers}, E.~A. {Magnier}, N.~{Metcalfe}, et~al., arXiv e-prints arXiv:1612.05560 (2016).

\bibitem[{Condon} et~al.(1998)]{1998AJ....115.1693C}
J.~J. {Condon}, W.~D. {Cotton}, E.~W. {Greisen}, et~al., \aj \textbf{115}~(5), 1693 (1998).

\bibitem[{Cutri} et~al.(2013)]{2013wise.rept....1C}
R.~M. {Cutri}, E.~L. {Wright}, T.~{Conrow}, et~al., {Explanatory Supplement to the AllWISE Data Release Products}, Explanatory Supplement to the AllWISE Data Release Products, by R. M. Cutri et al. (2013).

\bibitem[{Dabhade} et~al.(2017)]{2017MNRAS.469.2886D}
P.~{Dabhade}, M.~{Gaikwad}, J.~{Bagchi}, et~al., \mnras \textbf{469}~(3), 2886 (2017).

\bibitem[{Dabhade} et~al.(2020)]{2020A&A...642A.153D}
P.~{Dabhade}, M.~{Mahato}, J.~{Bagchi}, et~al., \aap \textbf{642}, A153 (2020).

\bibitem[{Dennett-Thorpe} et~al.(2002)]{2002MNRAS.330..609D}
J.~{Dennett-Thorpe}, P.~A.~G. {Scheuer}, R.~A. {Laing}, et~al., \mnras \textbf{330}~(3), 609 (2002).

\bibitem[{Denney} et~al.(2014)]{2014ApJ...796..134D}
K.~D. {Denney}, G.~{De Rosa}, K.~{Croxall}, et~al., \apj \textbf{796}~(2), 134 (2014).

\bibitem[{Dey} et~al.(2019)]{2019AJ....157..168D}
A.~{Dey}, D.~J. {Schlegel}, D.~{Lang}, et~al., \aj \textbf{157}~(5), 168 (2019).

\bibitem[{En{\ss}lin} and {Gopal-Krishna}(2001)]{2001ASPC..250..454E}
T.~A. {En{\ss}lin} and {Gopal-Krishna}, in R.~A. {Laing} and K.~M. {Blundell} (eds.), \emph{Particles and Fields in Radio Galaxies Conference}, \emph{Astronomical Society of the Pacific Conference Series}, vol. 250, p. 454 (2001).

\bibitem[{Fanaroff} and {Riley}(1974)]{1974MNRAS.167P..31F}
B.~L. {Fanaroff} and J.~M. {Riley}, \mnras \textbf{167}, 31P (1974).

\bibitem[{Gaia Collaboration} et~al.(2018)]{2018A&A...616A...1G}
{Gaia Collaboration}, A.~G.~A. {Brown}, A.~{Vallenari}, et~al., \aap \textbf{616}, A1 (2018).

\bibitem[{Glikman} et~al.(2018)]{2018ApJ...861...37G}
E.~{Glikman}, M.~{Lacy}, S.~{LaMassa}, et~al., \apj \textbf{861}~(1), 37 (2018).

\bibitem[{Glikman} et~al.(2022)]{2022ApJ...934..119G}
E.~{Glikman}, M.~{Lacy}, S.~{LaMassa}, et~al., \apj \textbf{934}~(2), 119 (2022).

\bibitem[{Gopal-Krishna} et~al.(2012)]{2012RAA....12..127G}
{Gopal-Krishna}, P.~L. {Biermann}, L.~{\'A}. {Gergely}, and P.~J. {Wiita}, Research in Astronomy and Astrophysics \textbf{12}~(2), 127 (2012).

\bibitem[{Gordon} et~al.(2021)]{2021ApJS..255...30G}
Y.~A. {Gordon}, M.~M. {Boyce}, C.~P. {O'Dea}, et~al., \apjs \textbf{255}~(2), 30 (2021).

\bibitem[{Gregory} et~al.(1996)]{1996ApJS..103..427G}
P.~C. {Gregory}, W.~K. {Scott}, K.~{Douglas}, and J.~J. {Condon}, \apjs \textbf{103}, 427 (1996).

\bibitem[{Hardcastle} et~al.(2009)]{2009MNRAS.393.1041H}
M.~J. {Hardcastle}, C.~C. {Cheung}, I.~J. {Feain}, and {\L}.~{Stawarz}, \mnras \textbf{393}~(3), 1041 (2009).

\bibitem[{Helfand} et~al.(2015)]{2015ApJ...801...26H}
D.~J. {Helfand}, R.~L. {White}, and R.~H. {Becker}, \apj \textbf{801}~(1), 26 (2015).

\bibitem[{Helou} et~al.(1995)]{1995ASSL..203...95H}
G.~{Helou}, B.~F. {Madore}, M.~{Schmitz}, et~al., in D.~{Egret} and M.~A. {Albrecht} (eds.), \emph{Information \& On-Line Data in Astronomy}, vol. 203, p.~95 (1995).

\bibitem[{Hurley-Walker} et~al.(2017)]{2017MNRAS.464.1146H}
N.~{Hurley-Walker}, J.~R. {Callingham}, P.~J. {Hancock}, et~al., \mnras \textbf{464}~(1), 1146 (2017).

\bibitem[{Intema} et~al.(2017)]{2017A&A...598A..78I}
H.~T. {Intema}, P.~{Jagannathan}, K.~P. {Mooley}, and D.~A. {Frail}, \aap \textbf{598}, A78 (2017).

\bibitem[{Ivezi{\'c}} et~al.(2002)]{2002AJ....124.2364I}
{\v{Z}}.~{Ivezi{\'c}}, K.~{Menou}, G.~R. {Knapp}, et~al., \aj \textbf{124}~(5), 2364 (2002).

\bibitem[{Jamrozy} et~al.(2008)]{2008MNRAS.385.1286J}
M.~{Jamrozy}, C.~{Konar}, J.~{Machalski}, and D.~J. {Saikia}, \mnras \textbf{385}~(3), 1286 (2008).

\bibitem[{Joshi} et~al.(2019)]{2019ApJ...887..266J}
R.~{Joshi}, G.~{Krishna}, X.~{Yang}, et~al., \apj \textbf{887}~(2), 266 (2019).

\bibitem[{Kaiser} et~al.(1997)]{1997MNRAS.292..723K}
C.~R. {Kaiser}, J.~{Dennett-Thorpe}, and P.~{Alexander}, \mnras \textbf{292}~(3), 723 (1997).

\bibitem[{Kapahi}(1989)]{1989AJ.....97....1K}
V.~K. {Kapahi}, \aj \textbf{97}, 1 (1989).

\bibitem[{Kimball} and {Ivezi{\'c}}(2008)]{2008AJ....136..684K}
A.~E. {Kimball} and {\v{Z}}.~{Ivezi{\'c}}, \aj \textbf{136}~(2), 684 (2008).

\bibitem[{Komberg} and {Pashchenko}(2009)]{2009ARep...53.1086K}
B.~V. {Komberg} and I.~N. {Pashchenko}, Astronomy Reports \textbf{53}~(12), 1086 (2009).

\bibitem[{Kronberg}(1994)]{1994RPPh...57..325K}
P.~P. {Kronberg}, Reports on Progress in Physics \textbf{57}~(4), 325 (1994).

\bibitem[{Kronberg} et~al.(2004)]{2004ApJ...604L..77K}
P.~P. {Kronberg}, S.~A. {Colgate}, H.~{Li}, and Q.~W. {Dufton}, \apjl \textbf{604}~(2), L77 (2004).

\bibitem[{Kronberg} et~al.(2001)]{2001ApJ...560..178K}
P.~P. {Kronberg}, Q.~W. {Dufton}, H.~{Li}, and S.~A. {Colgate}, \apj \textbf{560}~(1), 178 (2001).

\bibitem[{Ku{\'z}micz} and {Jamrozy}(2012)]{2012MNRAS.426..851K}
A.~{Ku{\'z}micz} and M.~{Jamrozy}, \mnras \textbf{426}~(2), 851 (2012).

\bibitem[{Ku{\'z}micz} et~al.(2018)]{2018ApJS..238....9K}
A.~{Ku{\'z}micz}, M.~{Jamrozy}, K.~{Bronarska}, et~al., \apjs \textbf{238}~(1), 9 (2018).

\bibitem[{Lara} et~al.(2001{\natexlab{a}})]{2001A&A...370..409L}
L.~{Lara}, W.~D. {Cotton}, L.~{Feretti}, et~al., \aap \textbf{370}, 409 (2001{\natexlab{a}}).

\bibitem[{Lara} et~al.(2001{\natexlab{b}})]{2001A&A...378..826L}
L.~{Lara}, I.~{M{\'a}rquez}, W.~D. {Cotton}, et~al., \aap \textbf{378}, 826 (2001{\natexlab{b}}).

\bibitem[{Lawrence} et~al.(2007)]{2007MNRAS.379.1599L}
A.~{Lawrence}, S.~J. {Warren}, O.~{Almaini}, et~al., \mnras \textbf{379}~(4), 1599 (2007).

\bibitem[{Liu}(2004)]{2004MNRAS.347.1357L}
F.~K. {Liu}, \mnras \textbf{347}~(4), 1357 (2004).

\bibitem[{Lucas} et~al.(2008)]{2008MNRAS.391..136L}
P.~W. {Lucas}, M.~G. {Hoare}, A.~{Longmore}, et~al., \mnras \textbf{391}~(1), 136 (2008).

\bibitem[{Machalski} et~al.(2001)]{2001A&A...371..445M}
J.~{Machalski}, M.~{Jamrozy}, and S.~{Zola}, \aap \textbf{371}, 445 (2001).

\bibitem[{Machalski} et~al.(2006)]{2006A&A...454...85M}
J.~{Machalski}, M.~{Jamrozy}, S.~{Zola}, and D.~{Koziel}, \aap \textbf{454}~(1), 85 (2006).

\bibitem[{Malarecki} et~al.(2015)]{2015MNRAS.449..955M}
J.~M. {Malarecki}, D.~H. {Jones}, L.~{Saripalli}, et~al., \mnras \textbf{449}~(1), 955 (2015).

\bibitem[{Matt} et~al.(2003)]{2003MNRAS.342..422M}
G.~{Matt}, M.~{Guainazzi}, and R.~{Maiolino}, \mnras \textbf{342}~(2), 422 (2003).

\bibitem[{Mauch} et~al.(2003)]{2003MNRAS.342.1117M}
T.~{Mauch}, T.~{Murphy}, H.~J. {Buttery}, et~al., \mnras \textbf{342}~(4), 1117 (2003).

\bibitem[{McConnell} et~al.(2020)]{2020PASA...37...48M}
D.~{McConnell}, C.~L. {Hale}, E.~{Lenc}, et~al., \pasa \textbf{37}, e048 (2020).

\bibitem[{Missaglia} et~al.(2019)]{2019A&A...626A...8M}
V.~{Missaglia}, F.~{Massaro}, A.~{Capetti}, et~al., \aap \textbf{626}, A8 (2019).

\bibitem[{Mostert} et~al.(2024)]{2024A&A...691A.185M}
R.~I.~J. {Mostert}, M.~S.~S.~L. {Oei}, B.~{Barkus}, et~al., \aap \textbf{691}, A185 (2024).

\bibitem[{Murgia}(2003)]{2003PASA...20...19M}
M.~{Murgia}, \pasa \textbf{20}~(1), 19 (2003).

\bibitem[{Murgia} et~al.(1999)]{1999A&A...345..769M}
M.~{Murgia}, C.~{Fanti}, R.~{Fanti}, et~al., \aap \textbf{345}, 769 (1999).

\bibitem[{Murgia} et~al.(2011)]{2011A&A...526A.148M}
M.~{Murgia}, P.~{Parma}, K.~H. {Mack}, et~al., \aap \textbf{526}, A148 (2011).

\bibitem[{Murthy} et~al.(1994)]{Murthy1994}
S.~K. {Murthy}, S.~{Kasif}, and S.~{Salzberg}, arXiv e-prints cs/9408103 (1994).

\bibitem[{Ochsenbein} et~al.(2000)]{2000A&AS..143...23O}
F.~{Ochsenbein}, P.~{Bauer}, and J.~{Marcout}, \aaps \textbf{143}, 23 (2000).

\bibitem[{Oei} et~al.(2023)]{2023A&A...672A.163O}
M.~S.~S.~L. {Oei}, R.~J. {van Weeren}, A.~R.~D.~J.~G.~I.~B. {Gast}, et~al., \aap \textbf{672}, A163 (2023).

\bibitem[{Oei} et~al.(2022)]{2022A&A...660A...2O}
M.~S.~S.~L. {Oei}, R.~J. {van Weeren}, M.~J. {Hardcastle}, et~al., \aap \textbf{660}, A2 (2022).

\bibitem[{Olsen} et~al.(2019)]{2019BAAS...51g..61O}
K.~{Olsen}, A.~{Bolton}, S.~{Juneau}, et~al., in \emph{Bulletin of the American Astronomical Society}, vol.~51, p.~61 (2019).

\bibitem[{Owen} and {Ledlow}(1994)]{1994ASPC...54..319O}
F.~N. {Owen} and M.~J. {Ledlow}, in G.~V. {Bicknell}, M.~A. {Dopita}, and P.~J. {Quinn} (eds.), \emph{The Physics of Active Galaxies}, \emph{Astronomical Society of the Pacific Conference Series}, vol.~54, p. 319 (1994).

\bibitem[{Owen} and {Rudnick}(1976)]{1976ApJ...205L...1O}
F.~N. {Owen} and L.~{Rudnick}, \apjl \textbf{205}, L1 (1976).

\bibitem[{Panessa} et~al.(2019)]{2019NatAs...3..387P}
F.~{Panessa}, R.~D. {Baldi}, A.~{Laor}, et~al., Nature Astronomy \textbf{3}, 387 (2019).

\bibitem[{Parma} et~al.(1999)]{1999A&A...344....7P}
P.~{Parma}, M.~{Murgia}, R.~{Morganti}, et~al., \aap \textbf{344}, 7 (1999).

\bibitem[{Peng} et~al.(2015)]{2015aska.confE.109P}
B.~{Peng}, R.~R. {Chen}, and R.~{Strom}, in \emph{Advancing Astrophysics with the Square Kilometre Array (AASKA14)}, p. 109 (2015).

\bibitem[{Pirya} et~al.(2012)]{2012MNRAS.426..758P}
A.~{Pirya}, D.~J. {Saikia}, M.~{Singh}, and H.~C. {Chandola}, \mnras \textbf{426}~(1), 758 (2012).

\bibitem[{Planck Collaboration} et~al.(2020)]{2020A&A...641A...6P}
{Planck Collaboration}, N.~{Aghanim}, Y.~{Akrami}, et~al., \aap \textbf{641}, A6 (2020).

\bibitem[{Proctor}(2011)]{2011ApJS..194...31P}
D.~D. {Proctor}, \apjs \textbf{194}~(2), 31 (2011).

\bibitem[{Proctor}(2016)]{2016ApJS..224...18P}
D.~D. {Proctor}, \apjs \textbf{224}~(2), 18 (2016).

\bibitem[{Rengelink} et~al.(1997)]{Rengelink1997}
R.~B. {Rengelink}, Y.~{Tang}, A.~G. {de Bruyn}, et~al., Astronomy\&Astrophysics Suppl. Ser. \textbf{124}, 259 (1997).

\bibitem[{R{\"o}ttgering} et~al.(2011)]{2011JApA...32..557R}
H.~{R{\"o}ttgering}, J.~{Afonso}, P.~{Barthel}, et~al., Journal of Astrophysics and Astronomy \textbf{32}, 557 (2011).

\bibitem[{Safouris} et~al.(2009)]{2009MNRAS.393....2S}
V.~{Safouris}, R.~{Subrahmanyan}, G.~V. {Bicknell}, and L.~{Saripalli}, \mnras \textbf{393}~(1), 2 (2009).

\bibitem[{Saripalli} et~al.(2005)]{2005AJ....130..896S}
L.~{Saripalli}, R.~W. {Hunstead}, R.~{Subrahmanyan}, and E.~{Boyce}, \aj \textbf{130}~(3), 896 (2005).

\bibitem[{Schoenmakers} et~al.(2001)]{2001A&A...374..861S}
A.~P. {Schoenmakers}, A.~G. {de Bruyn}, H.~J.~A. {R{\"o}ttgering}, and H.~{van der Laan}, \aap \textbf{374}, 861 (2001).

\bibitem[{Solovyov} and {Verkhodanov}(2014{\natexlab{a}})]{2014AstBu..69..141S}
D.~I. {Solovyov} and O.~V. {Verkhodanov}, Astrophysical Bulletin \textbf{69}~(2), 141 (2014{\natexlab{a}}).

\bibitem[{Solovyov} and {Verkhodanov}(2014{\natexlab{b}})]{2014ARep...58..506S}
D.~I. {Solovyov} and O.~V. {Verkhodanov}, Astronomy Reports \textbf{58}~(8), 506 (2014{\natexlab{b}}).

\bibitem[{Subrahmanyan} et~al.(2008)]{2008ASPC..395..380S}
R.~{Subrahmanyan}, L.~{Saripalli}, V.~{Safouris}, and R.~W. {Hunstead}, in A.~H. {Bridle}, J.~J. {Condon}, and G.~C. {Hunt} (eds.), \emph{Frontiers of Astrophysics: A Celebration of NRAO's 50th Anniversary}, \emph{Astronomical Society of the Pacific Conference Series}, vol. 395, p. 380 (2008).

\bibitem[{Taylor}(2005)]{2005ASPC..347...29T}
M.~B. {Taylor}, in P.~{Shopbell}, M.~{Britton}, and R.~{Ebert} (eds.), \emph{Astronomical Data Analysis Software and Systems XIV}, \emph{Astronomical Society of the Pacific Conference Series}, vol. 347, p.~29 (2005).

\bibitem[{van Weeren} et~al.(2010)]{2010Sci...330..347V}
R.~J. {van Weeren}, H.~J.~A. {R{\"o}ttgering}, M.~{Br{\"u}ggen}, and M.~{Hoeft}, Science \textbf{330}~(6002), 347 (2010).

\bibitem[{Verkhodanov} et~al.(2016)]{2016AstBu..71..139V}
O.~V. {Verkhodanov}, D.~I. {Solovyov}, O.~S. {Ulakhovich}, and M.~L. {Khabibullina}, Astrophysical Bulletin \textbf{71}~(2), 139 (2016).

\bibitem[{Wenger} et~al.(2000)]{2000A&AS..143....9W}
M.~{Wenger}, F.~{Ochsenbein}, D.~{Egret}, et~al., \aaps \textbf{143}, 9 (2000).

\bibitem[{Willis} et~al.(1974)]{1974Natur.250..625W}
A.~G. {Willis}, R.~G. {Strom}, and A.~S. {Wilson}, \nat \textbf{250}~(5468), 625 (1974).

\bibitem[{Wright} et~al.(2010)]{2010AJ....140.1868W}
E.~L. {Wright}, P.~R.~M. {Eisenhardt}, A.~K. {Mainzer}, et~al., \aj \textbf{140}~(6), 1868 (2010).

\end{thebibliography}

\end{document}